\documentclass[amsmath, amssymb,10pt,aps,prb,twocolumn,notitlepage,showpacs,superscriptaddress]{revtex4-1}
\usepackage{graphicx}
\usepackage{calc}
\usepackage{braket}
\usepackage{ulem}   
\usepackage{slashed}
\usepackage{wasysym}
\usepackage{siunitx}
\usepackage{dsfont}
\usepackage{amsthm,amsmath,amsfonts,amssymb,verbatim,color}
\usepackage{graphicx}
\usepackage{subfigure}
\usepackage{bm}
\usepackage{epsfig,slashed}
\usepackage[T1]{fontenc}
\usepackage[colorlinks=true,citecolor=blue,linkcolor=blue,urlcolor=blue]{hyperref}
\usepackage[version=4]{mhchem}
\usepackage{gensymb}
\begin{document}
\title{Quantized Magneto-Terahertz Effects in the Antiferromagnetic Topological Insulator MnBi$_2$Te$_4$ Thin Films}

\author{Xingyue Han}
\affiliation{Department of Physics and Astronomy, University of Pennsylvania, Philadelphia, Pennsylvania 19104, USA}
\author{An-Hsi Chen}
\affiliation{Materials Science and Technology Division, Oak Ridge National Laboratory, Oak Ridge, Tennessee 37831, USA}
\author{Matthew Brahlek}
\affiliation{Materials Science and Technology Division, Oak Ridge National Laboratory, Oak Ridge, Tennessee 37831, USA}
\author{Liang Wu}
\email{liangwu@sas.upenn.edu}
\affiliation{Department of Physics and Astronomy, University of Pennsylvania, Philadelphia, Pennsylvania 19104, USA}

\date{\today}

\begin{abstract}

MnBi$_2$Te$_4$ (MBT) is an ideal platform for studying the interplay between magnetism and topology. Many exotic topological phenomena, such as the quantum anomalous Hall effect and the axion insulator, have been observed in few-layer MBT. A key feature in MBT is the emergence of the surface exchange gap, which lies in the milli-electron-volt range (1 THz corresponds to 4.14 meV). This makes THz spectroscopy a powerful tool to probe the associated topological physics. In this study, we report the THz spectra of Faraday and Kerr rotations in MBT thin films grown by molecular beam epitaxy. By varying the external magnetic field, we observe three magnetic states: the antiferromagnetic state, the canted antiferromagnetic state, and the ferromagnetic state. Our terahertz results show a quantized Hall state under 6 T in both 6 SL and 7 SL samples without gate voltage. These findings provide new insights into the magneto-terahertz properties of MBT and its potential for topological spintronic applications.

\end{abstract}

\maketitle

\textbf{Introduction}
\\
Topological insulators (TIs) are characterized by an insulating bulk and conducting surface states \cite{hasan2010rmp}. In the bulk, band inversion between the conduction and valence bands originates from the spin-orbit coupling. On the surface, the $Z_2$ topological index gives rise to the gapless states with a linear dispersion near the degeneracy point, also known as the Dirac point. When time-reversal symmetry is broken by introducing magnetization to a TI, a mass gap forms in the surface states, creating a magnetic topological insulator \cite{tokura2019nrp, Chang2013sci}. If the Fermi energy is tuned within this surface mass gap, the system hosts a quantized Hall conductivity, $\sigma_{xy} = e^2/h$ with a vanishing longitudinal conductivity, $\sigma_{xx}$. For the magneto-optical properties, the system shows a quantized Faraday rotation ($\theta_F=\alpha \sim 7.3$ mrad, where $\alpha$ is the fine structure constant) and Kerr rotation ($\theta_K=\pi/2$ rad) from a free-standing film\cite{tse2010prl}. If the film is grown on a substrate, the quantized values of the optical rotations are modified by the refractive index of the substrate $n_s$ \cite{okada2016natcomm}. Magnetization in TIs can be introduced through three primary methods: doping with magnetic ions \cite{Chang2013sci, han2024nano, okada2016natcomm, mogi2015apl}, interfacing with magnetic insulators \cite{eremeev2013prb}, or engineering intrinsic magnetic TI systems \cite{li2019sciad, Deng2020Sci, li2019prx, bielinski2025natphy, ovchinnikov2021nano}. The third method is exemplified by the MnBi$_2$Te$_4$ (MBT) family of materials\cite{ Deng2020Sci, li2019prx, bielinski2025natphy, ovchinnikov2021nano}.

MBT is a van der Waals (vdW) layered material. Each vdW layer consists of a stacking  of Te-Bi-Te-Mn-Te-Bi-Te atomic layer, which is called a septuple layer (SL) \cite{gong2019cpl, otrokov2019nature, Deng2020Sci}. Monolayer MBT can be viewed as a Mn-Te bilayer intercalated into the center of a Bi$_2$Te$_3$ quintuple layer (QL). MBT exhibits A-type antiferromagnetic ordering, in which the magnetic moments of Mn atoms within the same SL couple ferromagnetically in the out-of-plane direction, while moments from adjacent SLs couple antiferromagnetically. This magnetic structure gives rise to a layer-dependent behavior: even-layer MBT has zero net magnetization in the bulk due to complete cancellation of opposite moments, whereas odd-layer MBT exhibits finite uncompensated magnetization in the bulk \cite{yang2021prx, zhao2021nano, li2024natcomm, Deng2020Sci}. The topology in MBT is similar to the nonmagnetic \ce{Bi2Te3}, where the $p_z$ bands from Bi and Te reverse under spin-orbit coupling \cite{li2019sciad, li2019prx}. The interplay between magnetism and topology makes MBT an ideal platform for exploring exotic topological states. For instance, in the transport measurements, the QAH effect has been observed in 5-SL MBT thin flakes at 6.5 K \cite{Deng2020Sci}. Similarly, the axion insulator state has been reported in 6-SL MBT thin flakes\cite{liu2020natmat}. Theoretical calculations also predict interesting magneto-optical effects \cite{Lei2023prb}. Without magnetic field, Faraday and Kerr rotations are predicted to be quantized ($\theta_F\sim 7.3$ mrad, $\theta_K=\pi/2$ rad) in 5, 7, 9 SL MBT, but vanishing in even-layers. With an out-of-plane magnetic field aligning the spins,  all thicknesses from 3 to 8 SL are predicted to show such quantization. However, the experimental investigations have been still lacking.

\begin{figure*}
\centering
\includegraphics[width=0.7\textwidth]{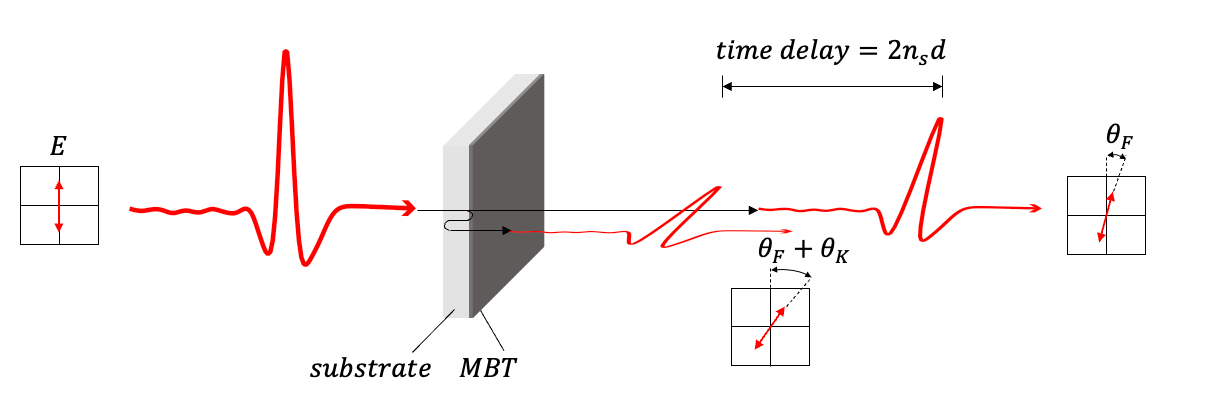}
\caption{Sketch of the time-domain THz setup. The incident THz light is polarized vertically. The transmitted light carries several pulses separated in the time domain. Here we only show the directly transmitted pulse and the first echo. The first pulse is the directly transmitted pulse. Its polarization is rotated by $\theta_F$, the Faraday rotation. The second pulse experiences an extra beam path in the substrate from the reflection. Its polarization is rotated by $\theta_F+\theta_K$, sum of the Faraday rotation and Kerr rotation. The time delay between the two pulses is $2n_sd$, where $n_s$ is the refractive index of the substrate, and $d$ is the thickness of the substrate.}
\label{Fig1}
\end{figure*}

The magnitude of the MBT surface gap varies in different reports, but it consistently falls within the milli-electron-volt range \cite{li2024nsr, otrokov2019nature, ji2021cpl, li2019prx, tan2023prl, ma2021prb}. This energy scale makes terahertz (THz) spectroscopy an ideal tool to investigate the topological phenomena associated with the surface gap, as 1 THz corresponds to 4.14 meV. Time-domain THz spectroscopy has been already proven powerful in studying cyclotron resonance\cite{wu2015prl} and the QAH state in magnetically modulated doped TIs \cite{okada2016natcomm}. However, THz studies on intrinsic magnetic TIs such as MBT remain scarce. In this work, we investigate MBT thin films grown by molecular beam epitaxy (MBE) using time-domain THz spectroscopy. By employing a polarization-sensitive technique \cite{han2024nano, Han2022PRB, okada2016natcomm, wu2016sci, dziom2017natcomm}, we examine the Faraday and Kerr rotations in MBT samples under a static external magnetic field up to 6 Tesla. This setup allows us to probe the THz response across different magnetic states of MBT. On a 6-SL sample, we observe both Faraday and Kerr rotations exhibiting abrupt magnitude changes during the transitions from the antiferromagnetic (AF) to canted-antiferromagnetic (cAF) state and from the cAF to ferromagnetic (FM) state. Furthermore, we observe quantized Hall conductance in both 6 SL and 7 SL sample under 6 T. These results shed new light on the magneto-terahertz characteristics of MBT and its promising role in topological quantum applications.

\textbf{Results}

The MBT films are grown by molecular beam epitaxy (MBE) on 4 mm $\times$ 4 mm (0001)-oriented Al$_2$O$_3$ substrates, with a protective 40 nm Te capping layer deposited in-situ \cite{lapano2020prm}. High-purity Bi, Mn, and Te sources are calibrated using an in situ quartz crystal microbalance (QCM) to achieve the desired flux ratios. The QCM measurements are cross-validated with film thicknesses determined via x-ray reflectivity, ensuring a flux ratio accuracy within a few percent. Thicknesses of the MBT films are characterized by x-ray reflectivity (see supplementary information). Given that the THz beam size is typically on the order of millimeter size restricted by the diffraction limit, it is essential to use MBE-grown films rather than exfoliated flakes to ensure the uniform interaction across the entire beam profile.

We employ time-domain THz polarimetry to investigate the magneto-terahertz Faraday and Kerr effects on MBT thin films. The sketch of the experimental setup is shown on FIG.\ref{Fig1}. Before the sample, the incident THz is vertically polarized by a wire-grid polarizer. Upon illumination by the linearly polarized THz light, the sample induces a change in the polarization plane of the transmitted light, which is analyzed by another wire-grid polarizer. The transmitted light contains a series of THz pulses due to the reflection at the substrate-film interface. The directly transmitted light (also known as the "main pulse") arrives at the detector first, and its polarization is rotated by $\theta_F$, the Faraday angle. The first echo arrives at the detector at a later time, and its polarization is rotated by $\theta_F+\theta_K$, the sum of the Faraday rotation and Kerr rotation. The reflection at the substrate-film interface gives rise to the $\theta_K$ rotation. Due to the additional light path, the echo is also separated from the main pulse in time domain by $2n_sd$, where $n_s$ is the refractive index of the substrate, and $d$ is the thickness of the substrate. The time-domain THz measurement has two key advantages. First, the technique enables simultaneous measurement of Faraday and Kerr rotations within a single scan \cite{han2024nano}. By setting the time window properly, we can choose to analyze the main pulse and the echo.  Second, the frequency range of THz spectroscopy (0.4–1.6 THz, corresponding to photon energies of 1.6–6.6 meV) lies below or around the magnetic mass gap of MBT films \cite{liu2022pnas}, making it an ideal probe for the surface gap and related phenomena. The Faraday effect and Kerr effect can serve as a probe of the topological magnetoelectric (TME) effect \cite{krichevtsov1996prl}. In TIs, the ME susceptibility is quantized in units of the fine structure constant $\alpha$ \cite{okada2016natcomm}. Theoretical calculations predict magneto-optical effect in MBT thin films and its thickness dependence: in antiferromagnetic state, both Faraday and Kerr effect vanish in even-number layer samples; in ferromagnetic state, Faraday and Kerr effect emerge in all samples, regardless of the thickness being even or odd \cite{Lei2023prb}. They have not yet to be experimentally investigated.

\begin{figure*}
\centering
\includegraphics[width=\textwidth]{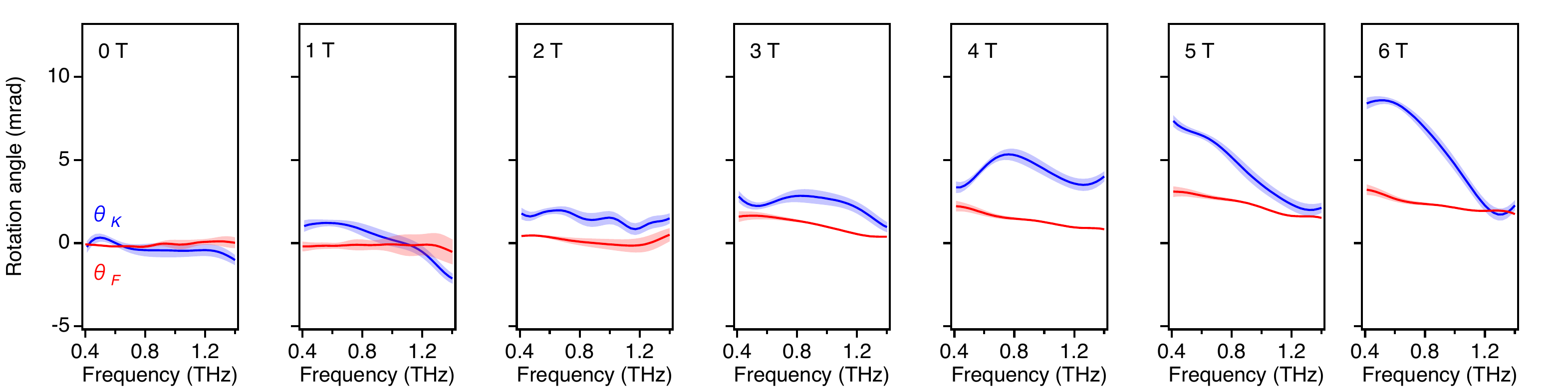}
\caption{THz Faraday and Kerr rotation spectra of the 6-SL MBT film at 3 K under varying magnetic fields. The shaded area indicates the error bar, defined as the standard deviation of three consecutive measurements.}
\label{Fig2}
\end{figure*}

\begin{figure*}
\centering
\includegraphics[width=\textwidth]{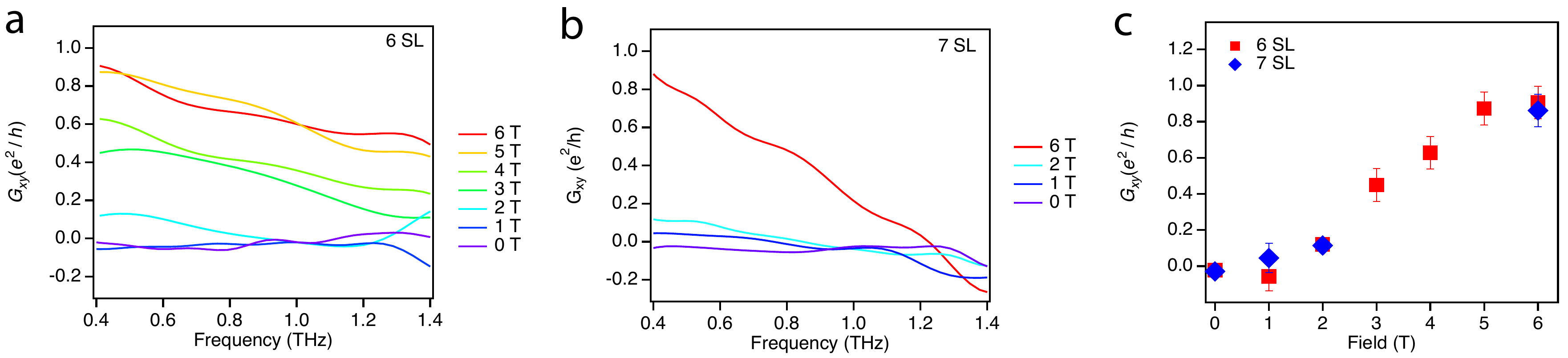}
\caption{\textbf{a.} Hall conductance spectrum of the 6-SL MBT film at 3 K. \textbf{b.} Hall conductance spectrum of the 7-SL MBT film at 3 K. \textbf{c.} Hall conductance at 0.4 THz under various magnetic fields. Error bar is the standard deviation of repeating measurements.}
\label{Fig4}
\end{figure*}

Here we present the Faraday and Kerr rotation spectra of the 6-SL MBT film at 3 K, well below the Néel temperature of 22 K \cite{zhao2021nano, yang2021prx}. We apply an external magnetic field, $B$, perpendicular to the film surface, varying from +6 T to -6 T. To eliminate nonmagnetic artifacts from the experimental setup, we symmetrize the results using $\theta_{F/K}(B) = (\theta_{F/K}(+B) - \theta_{F/K}(-B))/2$. The Faraday rotation spectra, $\theta_F(\omega)$, are shown in the red curves in FIG.\ref{Fig2}. When $B \leq 1$ T, $\theta_F$ is almost zero with negligible frequency and field dependence. At intermediate field strengths of 2–4 T, the sample exhibits a finite $\theta_F$, showing moderate frequency and field dependence. For example, under 3 T, at the lower frequency limit of 0.4 THz, $\theta_F$ reaches 1.6 $\pm$ 0.3 mrad, decreasing almost linearly with increasing frequency. The error bars in this work are defined by the standard deviation of repeating measurements at the same condition. At the upper frequency limit of 1.4 THz, $\theta_F$ falls below 0.39 $\pm$ 0.02 mrad. Overall, $\theta_F$ is slightly enhanced at 4 T. When $B$ increases to 5–6 T, $\theta_F$ saturates. This indicates that the magnetic moments in the 6-SL MBT film are fully aligned, forming a ferromagnetic (FM) state. The frequency dependence is similar to the 3–4 T regime, with $\theta_F$ decreasing almost linearly with frequency. At 0.4 THz, $\theta_F$ reaches 3.1 $\pm$ 0.3 mrad at 5 T and 3.2 $\pm$ 0.3 mrad at 6 T. These values approach the theoretical quantization of Faraday rotation in the thin-film limit: $\theta_{\mathrm{F}} = \tan^{-1}\left(\frac{2\alpha}{n_{\mathrm{s}}+1}\right) \approx 3.5$ mrad in the DC limit. 

The Kerr rotation spectra, $\theta_K(\omega)$, are shown in the blue curves in FIG.\ref{Fig2}.  In the absence of an external magnetic field, the $\theta_K$ spectrum is flat and smooth, remaining zero between 0.4 THz and 1.4 THz. When a magnetic field of 1 T is applied, a clear frequency dependence emerges, displaying a decreasing trend with increasing frequency. $\theta_K$ gradually increases as the magnetic field further increases. Under 6 T, the maximum $\theta_K$ in the low frequency reaches around 8.5 $\pm$ 0.3 mrad, slightly lower than the theoretical quantization of Kerr rotation in the thin-film limit: $\theta_{\mathrm{K}} = \tan^{-1}\left(\frac{4\alpha n_s}{n_{\mathrm{s}}^2 - 1}\right) \approx 10.5$ mrad. The deviation between the experimental data and the theoretical quantization values for both $\theta_F$ and $\theta_K$ arises because the latter are calculated based on the QAH conductivity at dc limit. These findings indicate the possible Chern insulator state in the 6-SL MBT film when the magnetic fields exceeds 5 T.

We then calculate the terahertz Hall conductance $G_{xy}$ of the 6 SL MBT films from the Faraday angle measurements. We use \cite{Han2022PRB}:

\[
G_{xy} = \frac{n_s + 1}{Z_0} \theta_F,
\]
where $Z_0$ is the vacuum impedance. The results are shown in \ref{Fig4}\textbf{a}. The field dependence of $G_{xy}$ at $\omega=0.4$ THz is summarized in FIG. \ref{Fig4}\textbf{c} with orange solid circle markers. Three stages of Hall conductance are revealed: 1. For $B \leq 1~\mathrm{T}$, $G_{xy}$ is small, approximately zero, corresponding to the AFM state with no net magnetization. 2. For $2~\mathrm{T} \leq B \leq 4~\mathrm{T}$, $G_{xy}$ exhibits a finite value which increases with increasing magnetic field but lower than quantization. 3. For $B \geq 5~\mathrm{T}$, $G_{xy}$ saturates around the 0.9 $e^2/h$, signifying the Chern insulator state. These results provide robust evidence for the topological nature of the FM state in 6-SL MBT films under high magnetic fields. Similar results are observed in another 7-SL film (fig.\ref{Fig4}\textbf{b,c}). $G_{xy}$ under 6 T can reach  0.91 $\pm$ 0.09 $e^2/h$ in 6 SL and 0.86 $\pm$ 0.09 $e^2/h$ in 7 SL at 0.4 THz, respectively. 


To summarize, we investigate the magneto-terahertz properties of MBT thin films using time-domain THz spectroscopy. They exhibit a magnetic field dependence from 0 T to 6 T. For the 6-SL and 7-SL MBT films, the Faraday and Kerr rotations reach values close to the quantization at high magnetic fields, indicating the realization of the Chern insulator state. These findings deepen our understanding of MBT as a rich platform for topology and magnetism and provide insight into its potential applications in topological spintronics.

\textbf{Data and materials availability:} All data needed to evaluate the conclusions are present in the paper. Additional data related to this paper could be requested from the authors.

\textbf{Acknowledgment:} This project at Penn is mainly sponsored by the Army Research Office and was accomplished under Grant Number W911NF-20-2-0166.  This work at ORNL was supported by the U. S. Department of Energy (DOE), Office of Science, Basic Energy Sciences (BES), Materials Sciences and Engineering Division.
\\


\textbf{Competing interests:} The authors declare that they have no competing interests. 

\bibliography{MBT}

\clearpage

\renewcommand{\thefigure}{SI\arabic{figure}}
\setcounter{figure}{0}
\section{Supplementary Information}
We use X-ray reflectivity (XRR) to characterize the thicknesses of the MBT films that were capped by Te at room temperature after the growth. XRR was performed using a 4-circle Panalytical MRD using Cu K$\alpha_1$ radiation. Under small incident angles, the reflectivity of thin films shows oscillations with the increasing $2\theta$ (Kiessig fringes) due to the interference off the interfaces. The short period oscillation ($\sim 0.3^\circ$) is that of the Te cap and the longer period is that of the film ($\sim 1^\circ$). We fit the X-ray reflectivity to extract the thicknesses of the two MBT films, which are 84 Å and 94 Å, respectively, and the Te capping layer is around 30 nm. The film thicknesses correspond to 5.8 SL and 7.1 SL, respectively. The error bar in the fitting and the roughness is $\pm 2$ Å, which is around 0.1-0.2 SL. We round the fit thicknesses in septuple layers (SL) to the nearest integer, which corresponds to 6 SL and 7 SL, respectively.

\begin{figure}
\centering
\includegraphics[width=0.5\textwidth]{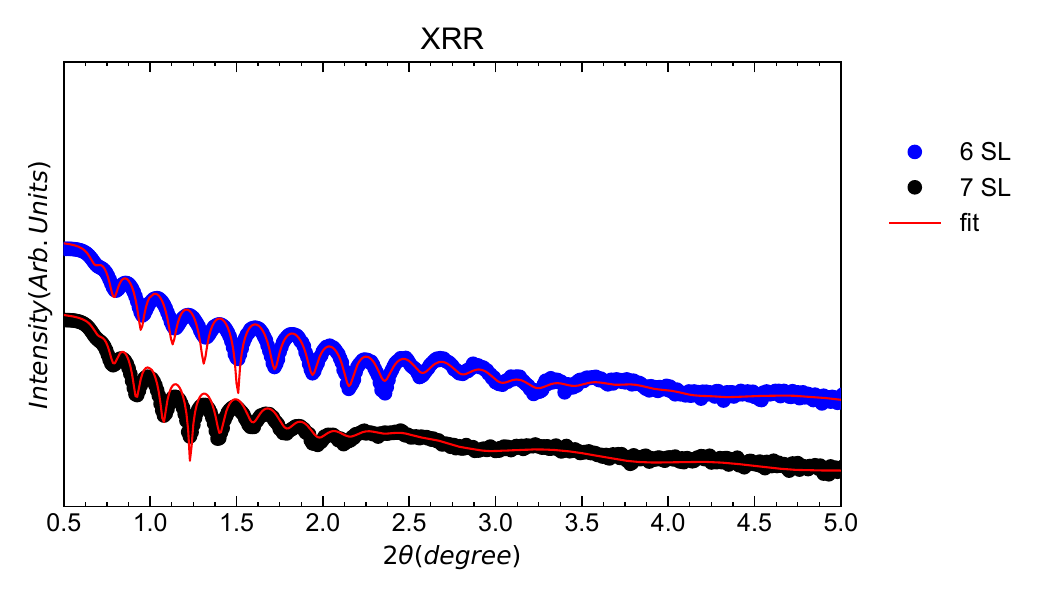}
\caption{X-ray reflectivity results on 6 SL and 7 SL MBT thin films.}
\label{FigSI}
\end{figure}

\end{document}